\documentclass[%
 preprint, 
 amsmath,amssymb,
 aps, physrev,
]{revtex4-2}

\usepackage{graphicx}
\usepackage{dcolumn}
\usepackage{bm}

\begin{document}

\preprint{APS/123-QED}

\title{\textbf{Surface scattering of atoms for high-sensitivity spectroscopy} 
}%

\author{V. J. Ajith}
 \author{Aaron Barr}
\author{Mark Raizen}%
 \email{Contact author: raizen@physics.utexas.edu}
\affiliation{%
 Authors' affiliations\\
  Department of Physics, The University of Texas at Austin, Austin, Texas, USA.
}%

\date{\today}

\begin{abstract}
High-sensitivity laser spectroscopy is integral to applications like atomic clocks, quantum computers, and chemical sensing. Lowering atomic temperature decreases spectral Doppler broadening and increases transit time across the excitation laser. We find that a polydimethylsiloxane (PDMS) coated surface can cool iron atoms from $\sim$1400 K to room temperature by surface scattering. It is also apparent that a single scattering event is enough for atoms to equilibrate with the PDMS surface temperature. Further, we show the very low adsorption of iron and ytterbium atoms on PDMS, an effect that persists to surface temperatures of 200 K. Through numerical simulation, we demonstrate the potential use of surface scattering in making a room temperature source of collimated atoms with enhanced flux and reduced velocity compared to that without surface scattering.  

\end{abstract}

\maketitle


\section{\label{sec:intro}Introduction}

Laser-based spectroscopic techniques are routinely used in laboratories for optical spectroscopy, which requires high sensitivity\cite{demtroder_applications_2015}. Atomic clocks\cite{ludlow_optical_2015,derevianko_colloquium_2011}, quantum computation\cite{ladd_quantum_2010}, atomic collision studies\cite{walker_measurements_1994,berman_study_1978, brewer1963study}, stable isotope  tracers\cite{kerstel_advances_2008}, chemical sensing\cite{harmon_laser-induced_2006,delucia_laser-induced_2005,wysocki_molecular_2010}, and monitoring of alloy compositions\cite{demtroder_applications_2015} are a few applications where high sensitivity laser spectroscopy plays an integral role. 

Doppler broadening is the main factor limiting the resolution of laser spectroscopy. This broadening is due to the distribution of atom velocity components along the laser direction, producing shifts in the resonant frequency of the atoms. The Doppler broadening can be reduced by collimating the atoms and exciting them with a laser perpendicular to the collimated atom beam. Collimation, typically with a slit or aperture, causes the atom beam to suffer a considerable loss of flux. Another way to mitigate Doppler broadening is to reduce the temperature. Temperature is typically fixed by the oven used to create the atom vapor. Reducing the oven temperature will reduce the flux and the signal. So, a separate way to cool the atoms while maintaining the vapor pressure is needed. 

The cooling and slowing of atoms not only helps in reducing Doppler broadening, it also increases the transit time of atoms across the excitation laser. The longer the transit time, the more excitation-emission cycles can occur in an atom, thus improving the signal from the atom. Atoms from an effusion oven have typical temperatures of 500 K to 2000 K. Atoms from an iron effusion oven\cite{huet_isotope_2015} operating at 1970 K have a most probable velocity ($v_{mp}$) of 758 m/s, while the same iron atoms at 298 K have $v_{mp}$ of 298 m/s, potentially doubling transit time.  

Atom slowing is a key step in optical trapping \cite{phillips_nobel_1998}. Trapped atoms are currently used in optical clocks\cite{ludlow_optical_2015}, quantum computing\cite{ladd_quantum_2010}, quantum sensing\cite{degen_quantum_2017}, studies of Bose-Einstein condensation\cite{anderson_observation_1995}, etc. Atoms from an oven are cooled to the K and mK ranges with techniques like Zeeman cooling\cite{phillips_nobel_1998}, chirp-cooling\cite{prodan_chirping_1984}, or buffer gas cooling\cite{hutzler_buffer_2012}. These techniques are used to slow atoms before trapping. Buffer gas cooling is a general technique, while Zeeman-cooling and chirp-cooling require significant changes for atoms of different elements. Buffer gas requires special considerations in flow rate such that a sufficient vacuum for the experiment is maintained. Buffer gas cooled atoms have different transit velocities compared to the average velocity from the temperature. 

In addition to cold atom beams, atoms in room temperature vapor cells are used for atom traps\cite{anderson_observation_1995, kock2016laser, yasuda2017laser}. The inner walls of vapor cells are coated with anti-relaxation coatings\cite{stephens_study_1994, goldenberg_atomic_1961,rahman1987rb} of long-chain polymers like paraffin and polydimethylsiloxane (PDMS), increasing the probability that the atom is trapped\cite{stephens_study_1994, yi2008method}. Moreover, such coatings preserve the spin polarization of atoms over many collisions. Only a few studies examine the number of scattering events required for the atom to adsorb on the walls and become lost, and studies on how many scattering events it takes for an atom to equilibrate to the surface temperature are similarly rare\cite{goldenberg_atomic_1961,stephens_study_1994,chi_advances_2020,sekiguchi_scattering_2018}. Studies of atom scattering from such coatings are instead focused on anti-relaxation properties\cite{wu_wall_2021, liberman1986relaxation, budker2005microwave}. 

A few publications estimate the adsorption energy of atoms on anti-relaxation coatings\cite{atutov_accurate_2015, bouchiat_relaxation_1966, zhao_method_2009, goldenberg_atomic_1961,brewer1963study,liberman1986relaxation,rahman1987rb,budker2005microwave,ulanski_measurement_2011,yi2008method}. It is found that the physical adsorption of rubidium atoms on paraffin has an adsorption energy of 0.1 eV\cite{bouchiat_relaxation_1966}. The shallow adsorption energy of atoms on these coatings\cite {atutov_accurate_2015,bouchiat_relaxation_1966,zhao_method_2009} facilitates desorption of atoms with low-intensity light. This effect is called light-induced atomic desorption (LIAD)\cite{gozzini_light-induced_1993,atutov_light-induced_1999,kock2016laser,yasuda2017laser}. Interestingly, these coatings have dwell times a few orders of magnitude larger than the expected values calculated from desorption energies\cite{atutov_accurate_2015,asakawa_measurement_2021, ulanski_measurement_2011}. The dwell time of rubidium on PDMS is around 22 $\mu$s\cite{atutov_accurate_2015}. Such long dwell times persist even at cold temperatures, for certain coatings. From room temperature to 123 K, rubidium atoms on a tetracontane (paraffin) coated surface have a dwell time of around 4 $\mu \text{s}$\cite{asakawa_measurement_2021}. 

This paper proposes a novel method to cool atoms from temperatures as high as 1400 K to room temperature, 298 K, through surface scattering of hot incoming atoms with a PDMS coating. The paper begins with the methods used in this study. Atom temperature is estimated using laser fluorescence spectroscopy. In the results section, we estimate the temperatures of the incoming and scattered atoms and establish that almost all atoms are scattered from the surface. Another result is that a single adsorption-desorption event from the PDMS coating appears sufficient for atoms to equilibrate with the surface temperature. Finally, we explore the feasibility of surface scattering as a room-temperature atom beam source through a numerical simulation, discuss the advantages of such a source, and conclude with future applications of this technique.

\section{Methods}
In this section, experiments and simulations involving collimated iron and ytterbium atoms scattered from a PDMS-coated surface are described. We begin with the coating procedure used to prepare PDMS surfaces, and then detail our atom sources, fluorescence laser spectroscopy methods, and thickness monitor measurements. Finally, we present the details of the numerical simulation.

\subsection{\label{sec:level2} Substrate Preparation}

The scattering surfaces were produced by spin-coating and subsequently curing a thin layer of Dow Sylgard 184 onto substrates made from microscope slides or INFICON thickness monitor crystals. Sylgard is a trade name for a self-curing formulation of the silicone polymer polydimethylsiloxane (PDMS). The viscosity of uncured PDMS is too great to achieve a uniform thin film with rotation speeds accessible to the spin-coater. However, diluting the polymer in a compatible solvent, such as toluene, allows for the controlled production of thin films by carefully adjusting the ratio of solvent to PDMS. For this study, liquid PDMS was diluted with toluene in a 10:1 solvent/polymer ratio. This solution was transferred to the substrate via micropipette and spun on a spin-coater at 3000 rpm for 1 minute. The solvent rapidly evaporates during the spinning process, leaving a visually uniform spot. Curing the film was accomplished by placing the coated substrate into a dish on a laboratory hotplate heated to 65 $^{\circ}$C for 1-2 hours. An optical image of PDMS coated glass surface is shown in our supplement for this paper \cite{supplemental}. Earlier studies using a similar protocol with the polymer PMMA measured the resulting film thickness at roughly 1 $\mu \text{m}$\cite{hall_spin_1998}. Quartz crystal monitor estimates of the thickness of the PDMS films we produced were consistent with this value. The surface morphology of thin PDMS films has been well characterized in prior studies \cite{bracic_preparation_2014}. 

\subsection{Experimental Setup}

Experiments were performed with an iron atom beam produced by laser ablation and an ytterbium atom beam from an effusion oven. The high melting point and low vapor pressure of iron\cite{huet_isotope_2015} make it a potentially interesting test case for novel cold atom beam techniques. Furthermore,  Ytterbium is included in this study partly because the relatively stable higher flux that can be achieved with an effusive oven was beneficial for some measurements. Details of the ytterbium oven are available in earlier publications\cite{bucay_surface_2019}. This section focuses on the atomic iron beam generation. 

We find that the laser ablation of iron oxide powder generates a flux of vaporized atomic iron. 
This experiment used iron (iii) oxide ($\text{Fe}_2\text{O}_3$) powder, with particle size $<$ 5 $\mu$m, obtained from Sigma Aldrich. Iron oxide nanopowder,  with particle size $<$ 50 nm, was found to result in less consistent flux of atomic iron than the micrometer-grained powder. 

Preparation of the iron powder involved placing a small amount into a 1" x 1/2" alumina crucible and compressing it to produce a uniform surface for exposure to the infrared ablation laser. The powder was pressed in the crucible using a flat-bottomed tamping rod with a diameter nearly equal to the interior diameter of the crucible. The collimating aperture was created by mounting a thin copper plate with two 1 mm diameter through-holes to the top of the crucible. 

\begin{figure*}
\includegraphics[width=1.0\textwidth]{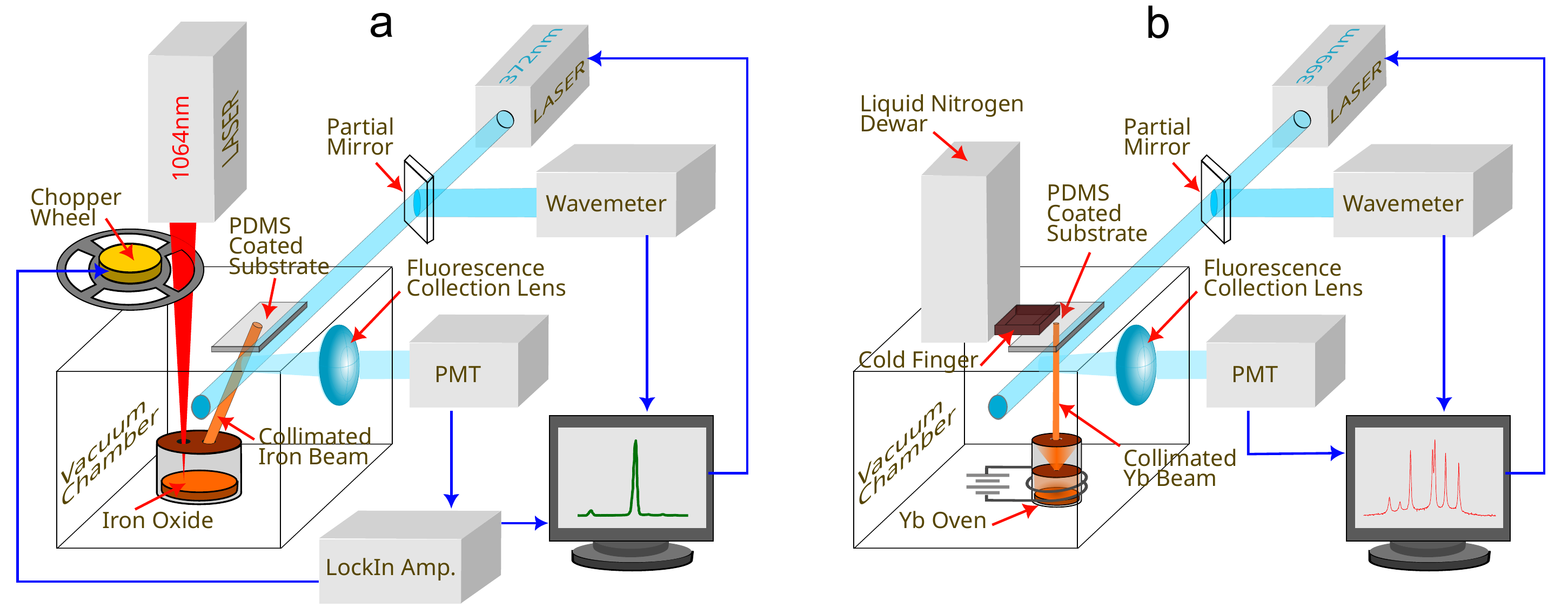}
\caption{\label{fig:schematics} Schematic figure of the experimental setup. Figures 'a' and 'b' show schematics of the setups used for experiments with iron and ytterbium atoms, respectively.}
\end{figure*}

\begin{figure}
\includegraphics[width=0.5\textwidth]{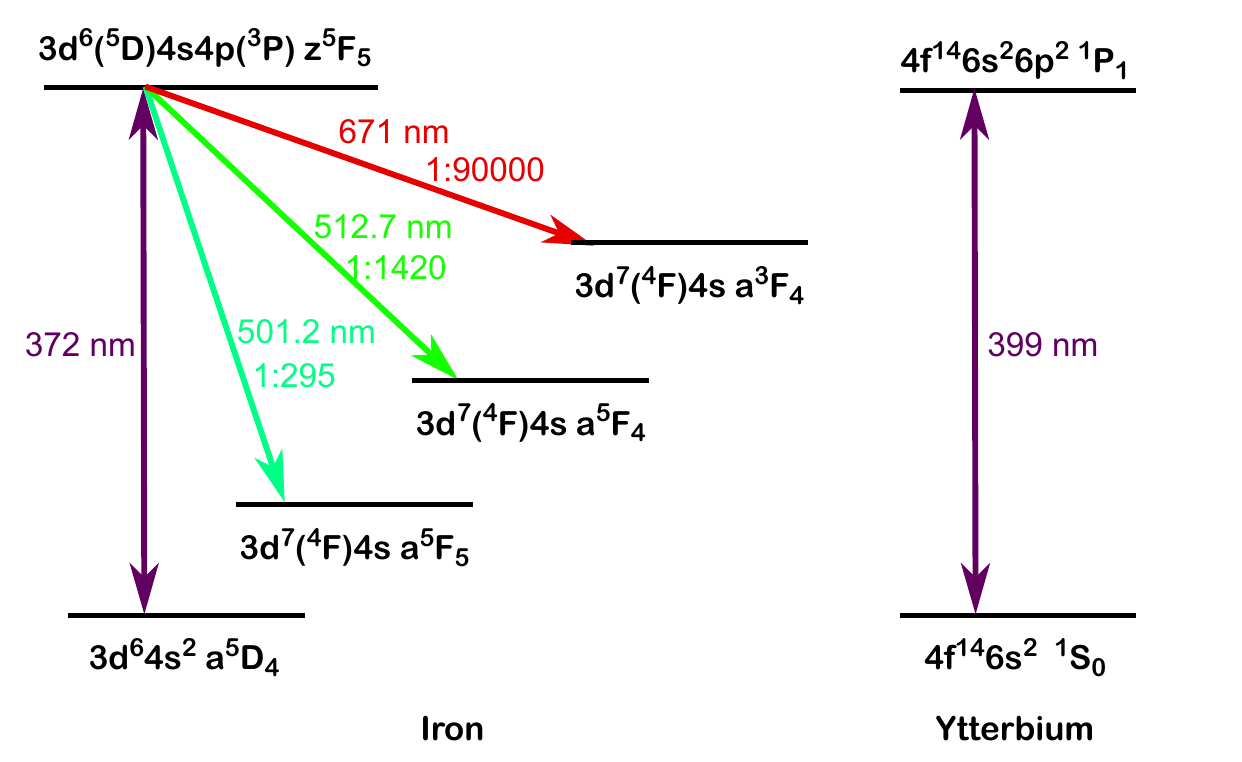}
\caption{\label{fig:level_dia} Grotrian diagram for 372 nm transition of atomic iron and 399 nm transition of atomic ytterbium (not to scale).}
\end{figure}

The schematic of the experimental setup is shown in the Fig.\ref{fig:schematics}. The sample-containing crucible was placed in vacuum, and then ablated using an IPG CW fiber-coupled 1064 nm YAG at 1-2 W. This ablation laser was chopped at 40 Hz, which was found to improve the stability of the fluorescence signal from the ablated iron. The IR ablation laser was focused on the powder surface through one of the two apertures in the top plate. The resulting collimated beam of atomic iron exiting the opposite aperture was slanted 15 $^0$ from the normal to the ablated powder surface. This slanted collimated beam was crossed at right angles by the CW fluorescence excitation laser, tuned to the 372 nm transition of iron (a$^5D_4$ $\leftrightarrow$ $z^5P_5$). The Grotrian diagram for this transition is shown in Fig.\ref{fig:level_dia}. The excitation laser was locked to a wavemeter, with computer control of the laser tuning parameters via a custom LabVIEW program. 

\begin{figure*}
\includegraphics[width=1\textwidth]{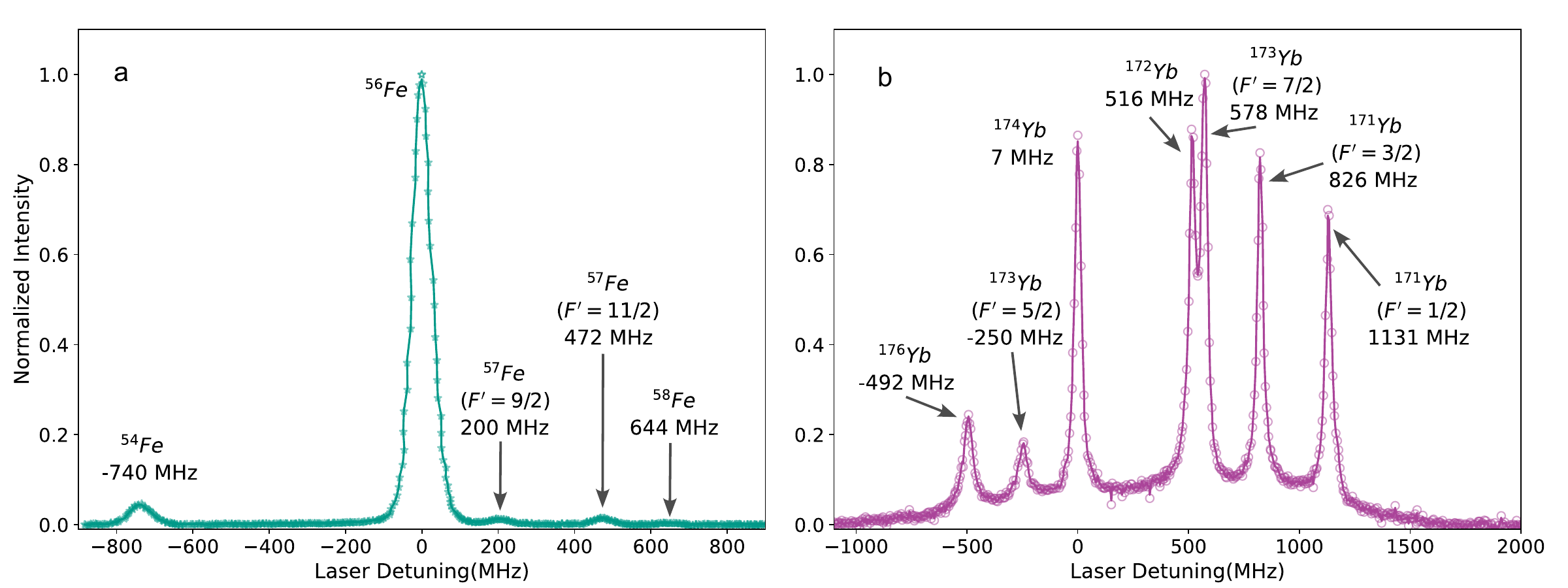}
\caption{\label{fig:spectrum} Fluorescence spectra of collimated iron (figure 'a') and ytterbium atoms (figure 'b') for the 372 nm and 399 nm transitions respectively. The spectra are normalized with respect to the highest peak intensity}
\end{figure*}

The resulting 372 nm fluorescence was collected and focused onto a photomultiplier tube (PMT), and the PMT signal was then fed into a lock-in amplifier using the chopper wheel as a reference. The excitation beam was tuned across a range of approximately 2 GHz while measuring the intensity of fluorescence, producing spectra in which peaks corresponding to each of the stable iron isotopes, $^{54}$Fe, $^{56}$Fe, $^{57}$Fe, and $^{58}$Fe, appear. Fig.\ref{fig:spectrum}a shows a typical iron spectrum. The heights of each peak provide a measure of relative isotopic abundance. 

A PDMS-coated substrate was attached to a feedthrough on the vacuum chamber and positioned just above the observation volume, where the excitation laser and collimated atom beam intersect, as shown in Fig.\ref{fig:schematics}. The collimated atom beam make a 15 $^0$ with respect to the PDMS surface normal and scatters from the PDMS surface. The scattered atoms are excited by the same laser along with the collimated atoms. 

This experiment was repeated with a beam of collimated ytterbium atoms incident normal to a PDMS-coated substrate, although without the flux modulation of the iron setup. Depending on the experiment, the substrate is mounted either on a liquid nitrogen cold finger or a quartz crystal thickness monitor. The ytterbium oven was run at 750 K and produced a constant flux throughout the experiment. We use the direct PMT voltage as a measure of fluorescence intensity.   

Spectra were produced for the 399 nm transition of ytterbium ($^1S_0$ $\leftrightarrow$ $^1P_1$). The Grotrian diagram for this transition is shown in Fig.\ref{fig:level_dia}. In Fig.\ref{fig:spectrum}b, the purple circles plot the spectrum of the collimated ytterbium atoms. The seven peaks resolved in the  spectra are $^{176}\text{Yb}$, $^{173}\text{Yb}$ $(\text{F}'=5/2)$, $^{174}\text{Yb}$, $^{172}\text{Yb}$, $^{173}\text{Yb}$ $(\text{F}'=7/2)$, $^{171}\text{Yb}$ $(\text{F}'=3/2)$ and $^{171}\text{Yb}$ $(\text{F}'=1/2)$, respectively.

For both iron and ytterbium experiments, the chamber pressure with the atom beam is around $10^{-6}$ to $10^{-5}$ Torr.

\subsection{Coating Thickness}

An INFICON quartz crystal monitor (QCM) was used to estimate the coating rate and deposited film thickness for an ytterbium atomic beam incident on both PDMS-coated and bare monitor crystals, which have a thin matte coating of gold. The mass change due to the deposition of ytterbium from the atomic beam, shifts the crystal's resonant frequency in a predictable way, providing a sensitive measurement of the coating thickness on the crystal. Normalized before-and-after measurements of the resonant frequency shift for both coated and bare crystals allow a direct comparison of the fraction of atoms scattered by coated and uncoated surfaces.

\subsection{Simulation of an atom beam source}

A simple numerical simulation was used to provide a proof-of-principle demonstration of polymer-coated surfaces as a cold atom source. Two cases were compared. In the first case, the flux and Doppler broadening of atoms incident on a collimating slit from a point source is considered. The second case introduces an intermediate scattering surface between the point source and the slit, examining the flux and Doppler broadening of atoms passing through the slit after scattering from the surface.

\begin{figure}
\includegraphics[width=0.5\textwidth]{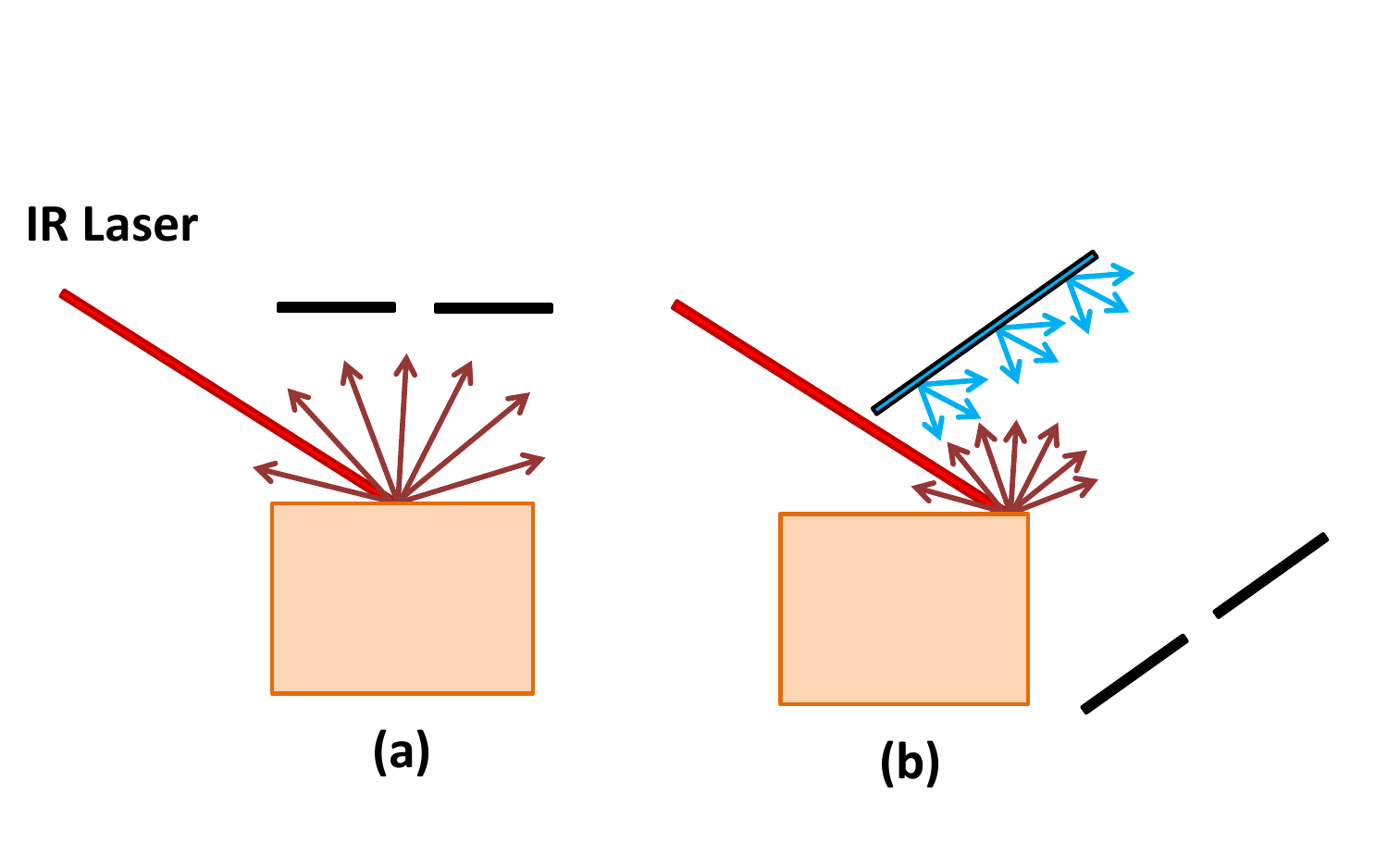}
\caption{\label{fig:sim_images} (a): Laser ablation and collimation by a slit. (b): Laser ablation followed by scattering from a surface, with subsequent collimation by a slit placed below the line of sight of the ablation point.}
\end{figure}

For the surface scattering model, a square surface with its midpoint lying some distance above the point of ablation scatters atoms towards a collimating slit. This slit is placed parallel to the scattering surface but away from the line of sight of the ablation point, located on the edge of a step, or "cliff," as shown in Fig. \ref{fig:sim_images}b.  The surface is angled such that part of the scattered flux reaches the slit, which is below the level of the step.

The flux from the ablation point is assumed to have a cosine distribution with respect to the surface normal \cite{sekiguchi_scattering_2018}. The atoms scattered from the surface are further assumed to thermally equilibrate with the surface, and scatter at room temperature. While the scattering surface is itself an extended source, since the scattering is approximated as purely diffuse, each individual point on the surface can be modeled as a point source of flux, with a cosine distribution relative to the surface normal . The behavior of the continuous extended source can be approximated with a sufficiently dense grid of scattering points; this study uses 10$^{6}$ points.  

The fraction of the original ablated flux which ultimately passes through the collimating slit can be computed by finding the fraction of the ablated flux which is scattered by the surface, and then the fraction of this scattered flux passing through the collimating slit.

\section{Results and Discussion}

\subsection{Temperature of Scattered Atoms}
The temperature of the collimated incoming atoms and the scattered atoms are estimated by fitting Voigt functions to the spectrum.

\begin{figure}
\includegraphics[width=0.5\textwidth]{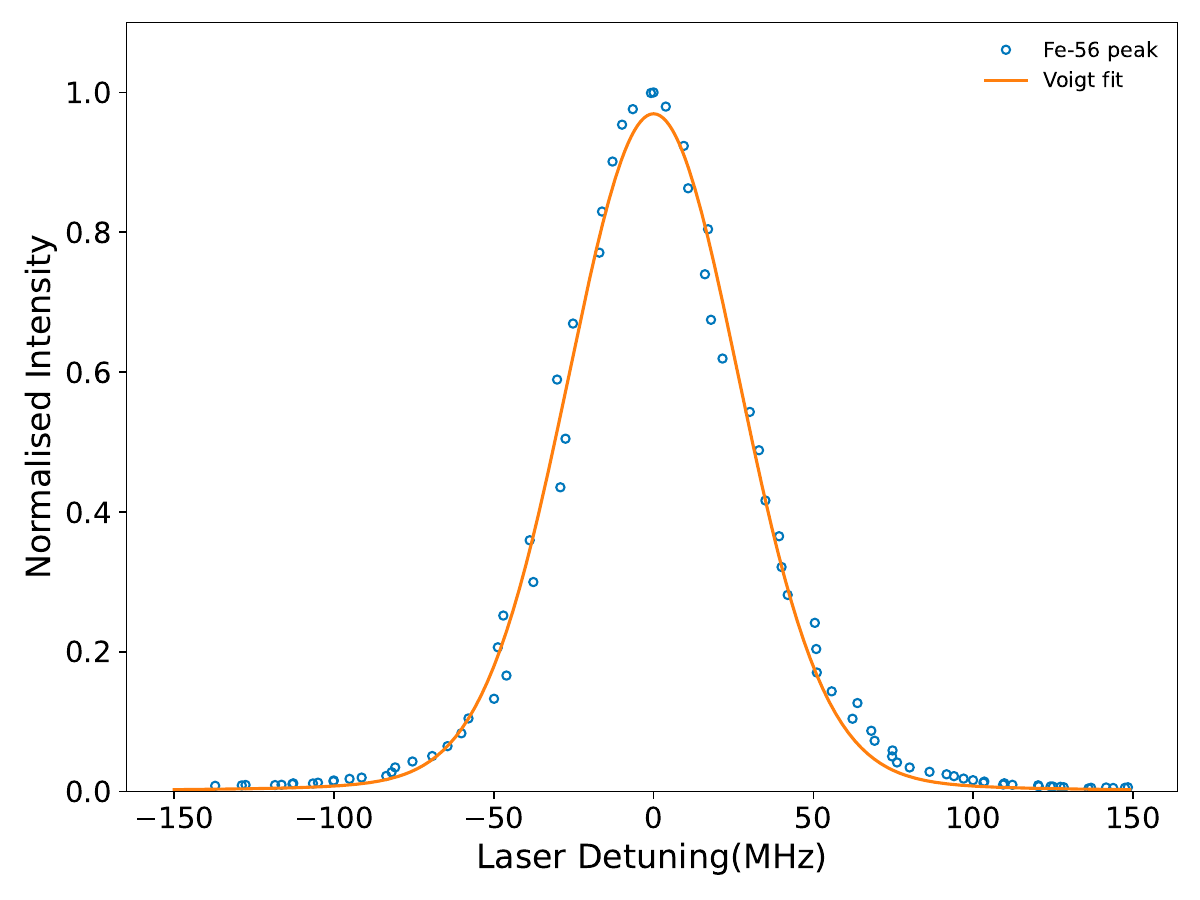}
\caption{\label{fig:incident_atoms} The temperature of the incident iron atoms is estimated by fitting a Voigt profile to the peak of an iron isotope in the spectrum of collimated atoms. Here the peak for $^{56}$Fe is fitted, obtaining a FWHM of 60 MHz. For a collimation ratio of 0.021, this FWHM corresponds to a temperature of 1418 K.}
\end{figure}

A spectrum from the iron atom beam for a collimation ratio of 0.021 is shown in Fig.\ref{fig:spectrum}. We define the collimation ratio as the radius of the collimating aperture divided by the distance the aperture have from the ablation point.   This spectrum shows the fluorescence peaks of the stable isotopes of iron for the 372 nm transition. We fit the isotope peaks with a Voigt function to estimate the temperature of atoms in the collimated beam. Fig.\ref{fig:incident_atoms} shows a representative fitting. The temperature estimate from this fit indicates that the incoming atoms are at a temperature of around 1418 $\pm$ 130 K. Uncertainity in the temperature comes from the uncertainity in the collimation ratio.

\begin{figure}
\includegraphics[width=0.5\textwidth]{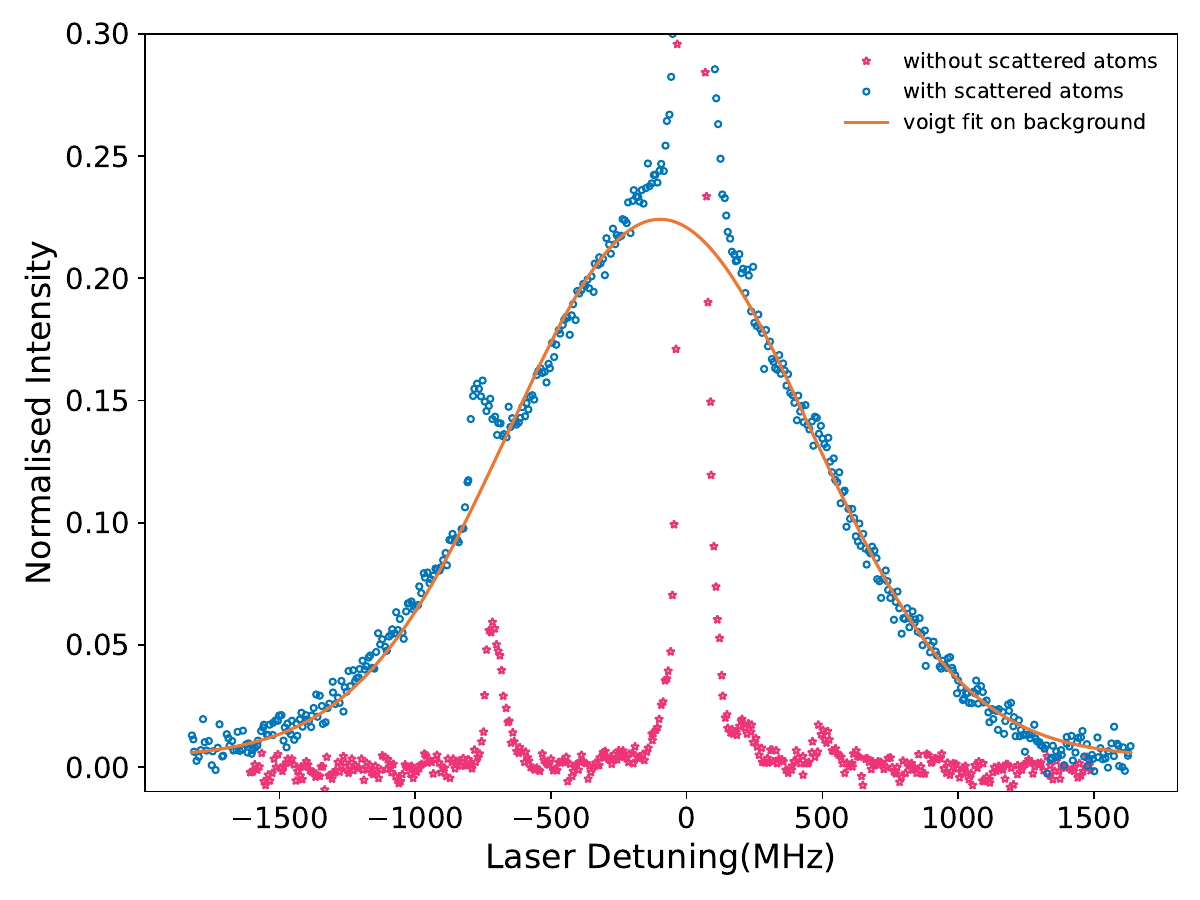}
\caption{\label{fig:spectrum_scattered_atoms} Fluorescence spectrum when the excitation laser is aligned very close to a PDMS coated glass surface, which scatters the collimated atom beam. The sharp peaks in the spectrum are due to the collimated atoms and the broad background is due to the scattered atoms. The blue circles show the data when a PDMS coated glass slide is used. For comparison, substrate-less spectrum with experimental parameters otherwise the same is plotted with pink stars. The background is fitted with a Voigt profile to estimate the temperature of the scatted atoms. Spectrum measured for a collimation ratio of 0.01 and lock-in time constant of 0.3 s}
\end{figure}

We then introduce a PDMS-coated glass surface into the path of this 1400 K iron atom beam. The 372 nm excitation laser is aligned adjacent to where the atom beam impacts the PDMS coating. Thus, both incoming collimated atoms and scattered atoms from the PDMS surface are excited by the 372 nm laser. We show the resulting spectrum in Fig.\ref{fig:spectrum_scattered_atoms}, with data points represented by blue circles. The same experimental parameters, but with the PDMS-coated glass surface removed, produce the spectrum with data points represented by pink stars. The spectrum with the PDMS-coated surface present has a background due to atoms scattered by the coating. This background is not due to fluorescence excited in the glass substrate or PDMS coating by the laser, as the lock-in amplifier would filter this background fluorescence. 

\begin{figure}
\includegraphics[width=0.5\textwidth]{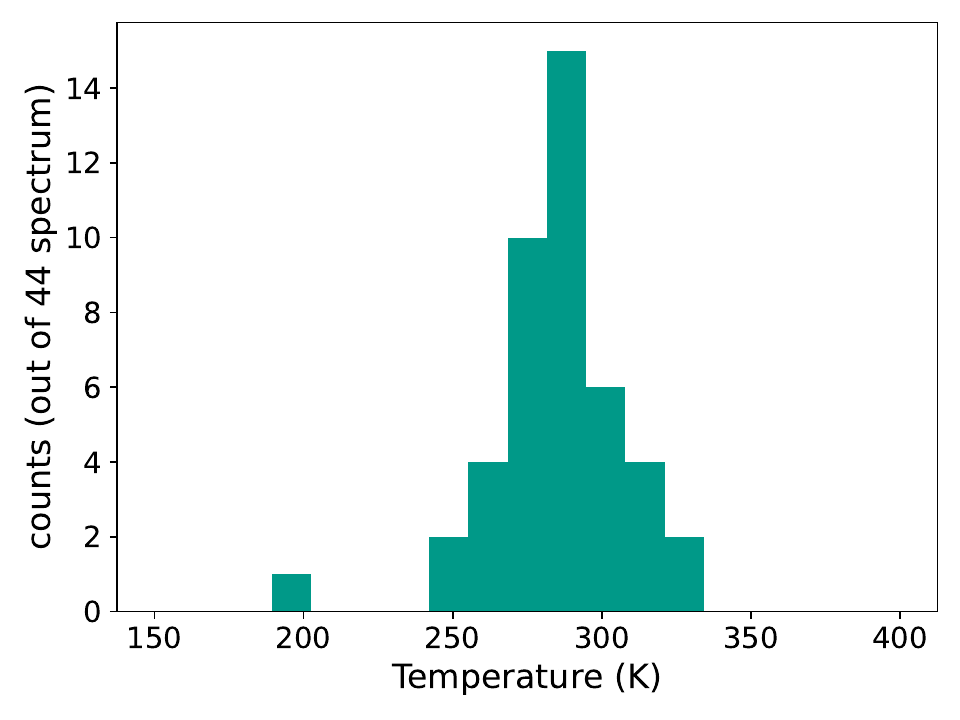}
\caption{\label{fig:hist} Histogram of the estimated temperatures from 44 independent spectra of scattered atoms from PDMS coated glass slides. The mean temperature of the scattered atoms is 285 K and the standard deviation is 22 K}
\end{figure}

The background spectrum is fitted with a Voigt function to extract the temperature of the scattered atoms. The solid orange line in Fig.\ref{fig:spectrum_scattered_atoms} shows a representative fit. The fit is repeated over multiple data sets as the substrate is moved to expose different areas of the PDMS coating to interact with the atom beam. The temperature estimates of the scattered atoms thus obtained from these fits are shown as a histogram in Fig.\ref{fig:hist}. We find that the mean temperature value of the scattered atoms is 285 K, and the standard deviation is 22 K. We expect the substrate to be equilibrated to the laboratory temperature, which is maintained at 293 K. Our measurement of the temperature of scattered atoms is close to the temperature of the surface from which the atoms scattered. This temperature, around 293 K, is lower than the original atom temperature of 1400 K by almost a factor of 5. The most probable velocity of the atoms is halved.  

The experiments suggest that the temperature equilibration occurs through a single scattering event. This is mainly because the collimated incoming atom beam is interacting with the PDMS surface only once, and outgoing scattered atoms move in straight lines away from the surface. We see fluorescence only in places where we expect atoms to move in straight-line paths; therefore, there is no build-up of iron or ytterbium atoms in the vacuum chamber. 

\subsection{Percentage of Scattered Atoms}

Next, a PDMS-coated surface was exposed to an atom beam flux for an extended period, and the degree of coating was measured. The laser ablation of iron powder does not yield the large stable iron atom flux needed to measure such a slow coating rate. An ytterbium beam produced by a hot oven\cite{bucay_surface_2019} was used instead. Exposing a thickness monitor crystal to the ytterbium atom beam for 100 minutes produced a visible coating of 162 nm. For a PDMS-coated surface, even at twice the exposure time (200 minutes), there is no visible coating on the crystal, and no increase in mass in the thickness monitor(QCM) measurements. This demonstrates that the majority of atoms interacting with the PDMS surface are scattered. Additionally, the high percentage of atoms scattered is not merely a transient behavior of freshly coated surfaces. The thickness monitor measurements were taken in the dark, eliminating the possibility of LIAD effects. The "non-stick" behavior of the PDMS coating is consistent with the low sticking probability of $\sim 10^{-5}$ measured for rubidium atoms on PDMS.\cite{atutov_accurate_2015}

\begin{figure}
\includegraphics[width=0.5\textwidth]{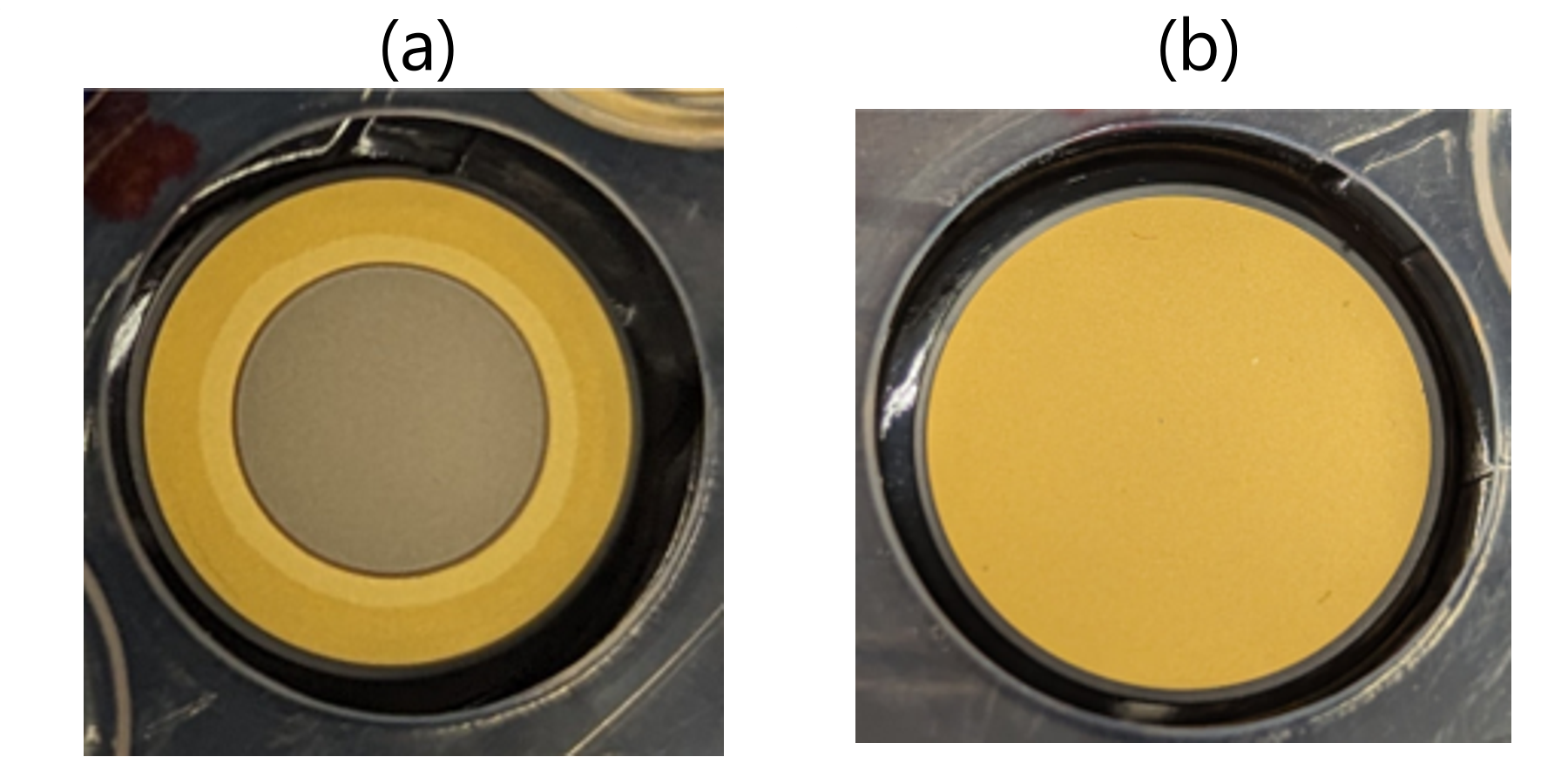}
\caption{\label{fig:thickness} Thickness monitor crystals (gold coated quartz) exposed to Yb atom beam. (a) Exposed to Yb atom beam for $\sim$100 minutes, resulting 162 nm Yb coating. (b) PDMS coated crystal exposed to similar flux of Yb atom beam for 200 minutes. No coating of Yb is visible or observed in measurements. }
\end{figure}

\subsection{\label{subsec:results:C}Substrate Cooling}

\begin{figure}
\includegraphics[width=0.5\textwidth]{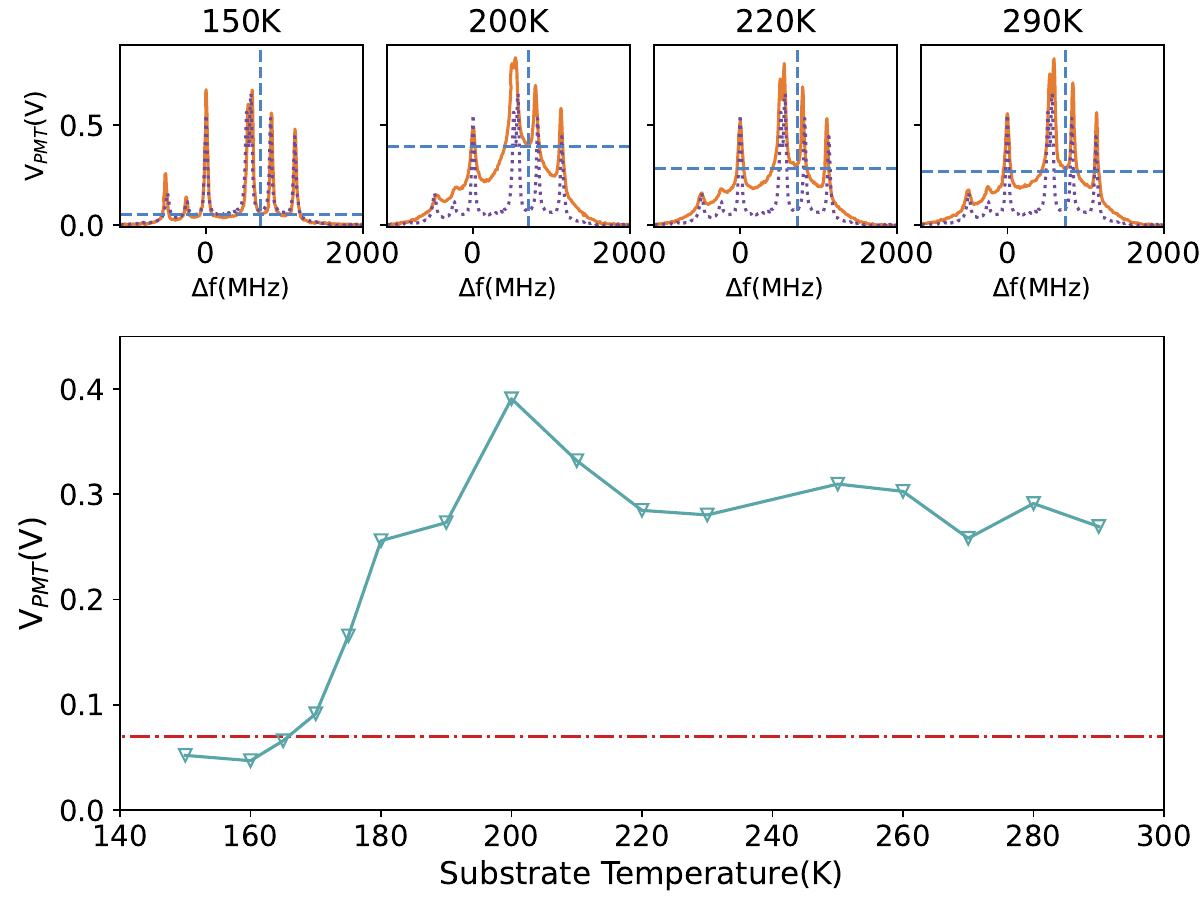}%
\caption{\label{fig:yb_liquid_n2}Top row shows ytterbium fluorescence spectra for 399 nm transition when the excitation laser is aligned very close to the surface of a PDMS coated thickness monitor crystal, which scatters the collimated atom beam. The sharp peaks in the spectrum are due to the collimated atoms and the broad background is due to the scattered atoms. Multiple sharp peaks in a given spectrum are because of the presence of different ytterbium isotopes in the collimated atom beam. Different spectra are for different substrate temperatures. For comparison, a substrate-less spectrum with the same experimental parameters is plotted with a purple dotted line. The blue dashed lines on each spectrum mark the local minima between the isotope peaks $^{173}\text{Yb}$ $(\text{F}'=7/2)$ and $^{171}\text{Yb}$ $(\text{F}'=3/2)$. This value is approximated as the maximum intensity of the background due to the scattered atoms and plotted separately (green triangles) in the bottom row for different substrate temperatures. The red "dot-dash" line shows the maximum intensity of the background for a substrate-less spectrum.}
\end{figure}

Next, we investigated whether ytterbium scattering persists in chilled substrates by mounting the substrate on a liquid nitrogen cold finger. Our cold finger takes roughly three hours to equilibrate to $\sim$ 100 K from room temperature. This cooling period is enough to take measurements at different surface temperatures. For every temperature, we detected the presence of atom scattering by observing the fluorescence spectrum of ytterbium. Like the iron spectrum, a broad background indicates the presence of scattered atoms. Unlike iron, ytterbium has more than one intense isotope peak in its spectrum. Thus, the background cannot be approximated by a single Voigt function, as was done for the iron spectrum. This non-trivial background fitting makes the temperature estimation of the scattered ytterbium atoms challenging. Nevertheless, the background of the spectrum indicates the presence of the scattered atoms.

The top row of Fig. \ref{fig:yb_liquid_n2} shows spectra for ytterbium when the excitation laser is aligned close to where the collimated atoms fall on a thickness monitor crystal coated with a 10\% PDMS solution. The y-axis is the PMT voltage without any normalization. Like the iron spectrum, the sharp peaks are due to the collimated incoming atoms, and the broad background is due to the scattered atoms. For reference, a spectrum when the substrate is removed from the path of incoming collimated atoms is plotted with a purple dotted line. The intensity maximum of the background is approximated by the local minima between the isotope peaks $^{173}\text{Yb}$ $(\text{F}'=7/2)$ and $^{171}\text{Yb}$ $(\text{F}'=3/2)$. In the representative spectra for a few substrate temperatures shown in the top row of Fig. \ref{fig:yb_liquid_n2}, the background intensity maxima are marked with dashed blue lines. The bottom row of Fig. \ref{fig:yb_liquid_n2} shows a plot of background intensity maxima versus substrate temperature. The difference between the background intensity maxima and the value for the substrate-less spectrum is proportional to the number of scattered atoms. Fig. \ref{fig:yb_liquid_n2} shows that the number of scattered atoms remains approximately unchanged for substrate temperatures in the range 290 K to 210 K, as the corresponding spectral backgrounds overlap, with similar background intensity maxima. Below 190 K, the background begins to decrease, and spectra below 170 K overlap the substrate-less spectrum. This trend is clearly seen in the background intensity maxima, where below 170 K, the values are comparable to the background intensity maxima of the substrate-less spectrum, shown with a "dot-dash" red line in Fig. \ref{fig:yb_liquid_n2}. This suggests that below substrate temperatures of 170 K, nearly all ytterbium atoms are being adsorbed on the PDMS surface. Changes in the physical properties of PDMS due to embrittlement below 200 K \cite{zhang_experimental_2020} could be the reason for this change in the scattering from the PDMS surface.

Although PDMS coatings do not scatter Yb atoms below 200 K, there could be other polymer coatings which can. In another study it is seen that the dwell time of rubidium on tetracontane is similar at both room temperature and 123 K\cite{asakawa_measurement_2021}. The same coating is found to scatter rubidium atoms at surface temperature\cite{sekiguchi_scattering_2018}. Tetracontane surfaces would be worth investigating for the lower-temperature scattering of atoms.

\subsection{\label{subsec:results:D}Atom Source}

In the numerical simulation, for purposes of comparison with the scattering case, a collimating slit 1 mm wide, 10 mm long, and lying 10 mm above the ablation point is considered. These values were chosen for ease of comparison. With these parameters, the slit produces a collimated atomic beam consisting of roughly 3.5\% of the original flux, with a Doppler width of around 50 MHz. 

For the scattering system, the scattering surface's height above the ablation point, its angle of tilt, and the dimensions of the surface all affect the Doppler broadening and flux of atoms passing through the slit (see Fig. \ref{fig:sim_images}b). If the surface is positioned close to the ablation point, it is possible to achieve an enhancement of flux relative to the non-scattering case. The enhancement calculation is shown in the supplemental materials \cite{supplemental}. Part of the flux that would otherwise be lost is scattered on a trajectory which passes through the slit. The degree of flux enhancement varies from a few percent to a maximum of $\sim$ 50\%.  

As an example, let us choose the surface to be an 8 mm x 8 mm square, with its midpoint 4 mm above the ablation point, and rotated nearly 45° with respect to the vertical. The angle of tilt is set to expose as much as possible of the surface to the ablated flux, while ensuring that every point on the surface has a line of sight to the collimating slit unobstructed by the "cliff" edge. The collimating slit is located 10 mm away from the surface, and is parallel to it. The slit length is again 10 mm, but the width is chosen to be 1.2 mm, to match the Doppler width of the slit-only case. 

For these parameters, the fraction of total flux through the collimating slit is around 5.2\%, with a Doppler width again of 50 MHz, meaning that the collimated flux scattered from the surface is improved by $\sim$ 50\% relative to the case for a slit alone, but at room temperature rather than 1400 K, and with the same degree of Doppler broadening.

\section{Conclusions}
We propose a novel method to cool atoms using atom-surface scattering. Our experiments show that $\sim$1 $\mu \text{m}$ coating of PDMS can scatter and cool $\sim$ 1400 K iron atoms to surface temperature. Further, we find that almost all ytterbium atoms falling on PDMS are scattered, and this scattering persists even for substrate temperatures as low as 200 K. Through a numerical simulation, we show the possibility of using the key findings of atomic thermal equilibration and "non-stick" properties of PDMS surfaces for a room temperature atom source. This simple technique helps improve the signal levels in high-sensitivity laser spectroscopy by reducing Doppler broadening without sacrificing flux, and by increasing the transit time of atoms in the experimental or observation volume. Moreover, the technique can be a first step in cooling atoms from high-temperature sources for cold atom experiments.

\begin{acknowledgments}
We wish to extend our sincere thanks to Patrick Obi, for his assistance with experiments. 

The authors gratefully acknowledge the support of the Copenhagen Center for Biomedical Quantum Sensing and the Pointsman Foundation.

\end{acknowledgments}

\section*{Data Availability Statement}

The data that support the findings of this study are available within the article.

\bibliography{ref_list}

\begin{thebibliography}{40}%
\makeatletter
\providecommand \@ifxundefined [1]{%
 \@ifx{#1\undefined}
}%
\providecommand \@ifnum [1]{%
 \ifnum #1\expandafter \@firstoftwo
 \else \expandafter \@secondoftwo
 \fi
}%
\providecommand \@ifx [1]{%
 \ifx #1\expandafter \@firstoftwo
 \else \expandafter \@secondoftwo
 \fi
}%
\providecommand \natexlab [1]{#1}%
\providecommand \enquote  [1]{``#1''}%
\providecommand \bibnamefont  [1]{#1}%
\providecommand \bibfnamefont [1]{#1}%
\providecommand \citenamefont [1]{#1}%
\providecommand \href@noop [0]{\@secondoftwo}%
\providecommand \href [0]{\begingroup \@sanitize@url \@href}%
\providecommand \@href[1]{\@@startlink{#1}\@@href}%
\providecommand \@@href[1]{\endgroup#1\@@endlink}%
\providecommand \@sanitize@url [0]{\catcode `\\12\catcode `\$12\catcode `\&12\catcode `\#12\catcode `\^12\catcode `\_12\catcode `\%12\relax}%
\providecommand \@@startlink[1]{}%
\providecommand \@@endlink[0]{}%
\providecommand \url  [0]{\begingroup\@sanitize@url \@url }%
\providecommand \@url [1]{\endgroup\@href {#1}{\urlprefix }}%
\providecommand \urlprefix  [0]{URL }%
\providecommand \Eprint [0]{\href }%
\providecommand \doibase [0]{https://doi.org/}%
\providecommand \selectlanguage [0]{\@gobble}%
\providecommand \bibinfo  [0]{\@secondoftwo}%
\providecommand \bibfield  [0]{\@secondoftwo}%
\providecommand \translation [1]{[#1]}%
\providecommand \BibitemOpen [0]{}%
\providecommand \bibitemStop [0]{}%
\providecommand \bibitemNoStop [0]{.\EOS\space}%
\providecommand \EOS [0]{\spacefactor3000\relax}%
\providecommand \BibitemShut  [1]{\csname bibitem#1\endcsname}%
\let\auto@bib@innerbib\@empty
\bibitem [{\citenamefont {Demtröder}(2015)}]{demtroder_applications_2015}%
  \BibitemOpen
  \bibfield  {author} {\bibinfo {author} {\bibfnamefont {W.}~\bibnamefont {Demtröder}},\ }\bibfield  {title} {\bibinfo {title} {Applications of {Laser} {Spectroscopy}},\ }in\ \href {https://doi.org/10.1007/978-3-662-44641-6_10} {\emph {\bibinfo {booktitle} {Laser {Spectroscopy} 2: {Experimental} {Techniques}}}}\ (\bibinfo  {publisher} {Springer Berlin Heidelberg},\ \bibinfo {year} {2015})\ pp.\ \bibinfo {pages} {589--650}\BibitemShut {NoStop}%
\bibitem [{\citenamefont {Ludlow}\ \emph {et~al.}(2015)\citenamefont {Ludlow}, \citenamefont {Boyd}, \citenamefont {Ye}, \citenamefont {Peik},\ and\ \citenamefont {Schmidt}}]{ludlow_optical_2015}%
  \BibitemOpen
  \bibfield  {author} {\bibinfo {author} {\bibfnamefont {A.~D.}\ \bibnamefont {Ludlow}}, \bibinfo {author} {\bibfnamefont {M.~M.}\ \bibnamefont {Boyd}}, \bibinfo {author} {\bibfnamefont {J.}~\bibnamefont {Ye}}, \bibinfo {author} {\bibfnamefont {E.}~\bibnamefont {Peik}},\ and\ \bibinfo {author} {\bibfnamefont {P.}~\bibnamefont {Schmidt}},\ }\bibfield  {title} {\bibinfo {title} {Optical atomic clocks},\ }\href {https://doi.org/10.1103/RevModPhys.87.637} {\bibfield  {journal} {\bibinfo  {journal} {Rev. Mod. Phys.}\ }\textbf {\bibinfo {volume} {87}},\ \bibinfo {pages} {637} (\bibinfo {year} {2015})}\BibitemShut {NoStop}%
\bibitem [{\citenamefont {Derevianko}\ and\ \citenamefont {Katori}(2011)}]{derevianko_colloquium_2011}%
  \BibitemOpen
  \bibfield  {author} {\bibinfo {author} {\bibfnamefont {A.}~\bibnamefont {Derevianko}}\ and\ \bibinfo {author} {\bibfnamefont {H.}~\bibnamefont {Katori}},\ }\bibfield  {title} {\bibinfo {title} {Colloquium: {Physics} of optical lattice clocks},\ }\href {https://doi.org/10.1103/RevModPhys.83.331} {\bibfield  {journal} {\bibinfo  {journal} {Rev. Mod. Phys.}\ }\textbf {\bibinfo {volume} {83}},\ \bibinfo {pages} {331} (\bibinfo {year} {2011})}\BibitemShut {NoStop}%
\bibitem [{\citenamefont {Ladd}\ \emph {et~al.}(2010)\citenamefont {Ladd}, \citenamefont {Jelezko}, \citenamefont {Laflamme}, \citenamefont {Nakamura}, \citenamefont {Monroe},\ and\ \citenamefont {O’Brien}}]{ladd_quantum_2010}%
  \BibitemOpen
  \bibfield  {author} {\bibinfo {author} {\bibfnamefont {T.~D.}\ \bibnamefont {Ladd}}, \bibinfo {author} {\bibfnamefont {F.}~\bibnamefont {Jelezko}}, \bibinfo {author} {\bibfnamefont {R.}~\bibnamefont {Laflamme}}, \bibinfo {author} {\bibfnamefont {Y.}~\bibnamefont {Nakamura}}, \bibinfo {author} {\bibfnamefont {C.}~\bibnamefont {Monroe}},\ and\ \bibinfo {author} {\bibfnamefont {J.~L.}\ \bibnamefont {O’Brien}},\ }\bibfield  {title} {\bibinfo {title} {Quantum computers},\ }\href {https://doi.org/10.1038/nature08812} {\bibfield  {journal} {\bibinfo  {journal} {Nature}\ }\textbf {\bibinfo {volume} {464}},\ \bibinfo {pages} {45} (\bibinfo {year} {2010})}\BibitemShut {NoStop}%
\bibitem [{\citenamefont {Walker}\ and\ \citenamefont {Feng}(1994)}]{walker_measurements_1994}%
  \BibitemOpen
  \bibfield  {author} {\bibinfo {author} {\bibfnamefont {T.}~\bibnamefont {Walker}}\ and\ \bibinfo {author} {\bibfnamefont {P.}~\bibnamefont {Feng}},\ }\bibfield  {title} {\bibinfo {title} {Measurements of {Collisions} {Between} {Laser}-{Cooled} {Atoms}},\ }in\ \href {https://doi.org/10.1016/S1049-250X(08)60076-2} {\emph {\bibinfo {booktitle} {Advances {In} {Atomic}, {Molecular}, and {Optical} {Physics}}}},\ Vol.~\bibinfo {volume} {34},\ \bibinfo {editor} {edited by\ \bibinfo {editor} {\bibfnamefont {B.}~\bibnamefont {Bederson}}\ and\ \bibinfo {editor} {\bibfnamefont {H.}~\bibnamefont {Walther}}}\ (\bibinfo  {publisher} {Academic Press},\ \bibinfo {year} {1994})\ pp.\ \bibinfo {pages} {125--170}\BibitemShut {NoStop}%
\bibitem [{\citenamefont {Berman}(1978)}]{berman_study_1978}%
  \BibitemOpen
  \bibfield  {author} {\bibinfo {author} {\bibfnamefont {P.~R.}\ \bibnamefont {Berman}},\ }\bibfield  {title} {\bibinfo {title} {Study of {Collisions} by {Laser} {Spectroscopy}},\ }in\ \href {https://doi.org/10.1016/S0065-2199(08)60055-X} {\emph {\bibinfo {booktitle} {Advances in {Atomic} and {Molecular} {Physics}}}},\ Vol.~\bibinfo {volume} {13},\ \bibinfo {editor} {edited by\ \bibinfo {editor} {\bibfnamefont {D.~R.}\ \bibnamefont {Bates}}\ and\ \bibinfo {editor} {\bibfnamefont {B.}~\bibnamefont {Bederson}}}\ (\bibinfo  {publisher} {Academic Press},\ \bibinfo {year} {1978})\ pp.\ \bibinfo {pages} {57--112}\BibitemShut {NoStop}%
\bibitem [{\citenamefont {Brewer}(1963)}]{brewer1963study}%
  \BibitemOpen
  \bibfield  {author} {\bibinfo {author} {\bibfnamefont {R.~G.}\ \bibnamefont {Brewer}},\ }\bibfield  {title} {\bibinfo {title} {Study of atom—wall collisions by optical pumping},\ }\href@noop {} {\bibfield  {journal} {\bibinfo  {journal} {The Journal of Chemical Physics}\ }\textbf {\bibinfo {volume} {38}},\ \bibinfo {pages} {3015} (\bibinfo {year} {1963})}\BibitemShut {NoStop}%
\bibitem [{\citenamefont {Kerstel}\ and\ \citenamefont {Gianfrani}(2008)}]{kerstel_advances_2008}%
  \BibitemOpen
  \bibfield  {author} {\bibinfo {author} {\bibfnamefont {E.}~\bibnamefont {Kerstel}}\ and\ \bibinfo {author} {\bibfnamefont {L.}~\bibnamefont {Gianfrani}},\ }\bibfield  {title} {\bibinfo {title} {Advances in laser-based isotope ratio measurements: selected applications},\ }\href {https://doi.org/10.1007/s00340-008-3128-x} {\bibfield  {journal} {\bibinfo  {journal} {Appl. Phys. B}\ }\textbf {\bibinfo {volume} {92}},\ \bibinfo {pages} {439} (\bibinfo {year} {2008})}\BibitemShut {NoStop}%
\bibitem [{\citenamefont {Harmon}\ \emph {et~al.}(2006)\citenamefont {Harmon}, \citenamefont {DeLucia}, \citenamefont {McManus}, \citenamefont {McMillan}, \citenamefont {Jenkins}, \citenamefont {Walsh},\ and\ \citenamefont {Miziolek}}]{harmon_laser-induced_2006}%
  \BibitemOpen
  \bibfield  {author} {\bibinfo {author} {\bibfnamefont {R.~S.}\ \bibnamefont {Harmon}}, \bibinfo {author} {\bibfnamefont {F.~C.}\ \bibnamefont {DeLucia}}, \bibinfo {author} {\bibfnamefont {C.~E.}\ \bibnamefont {McManus}}, \bibinfo {author} {\bibfnamefont {N.~J.}\ \bibnamefont {McMillan}}, \bibinfo {author} {\bibfnamefont {T.~F.}\ \bibnamefont {Jenkins}}, \bibinfo {author} {\bibfnamefont {M.~E.}\ \bibnamefont {Walsh}},\ and\ \bibinfo {author} {\bibfnamefont {A.}~\bibnamefont {Miziolek}},\ }\bibfield  {title} {\bibinfo {title} {Laser-induced breakdown spectroscopy – {An} emerging chemical sensor technology for real-time field-portable, geochemical, mineralogical, and environmental applications},\ }\href {https://doi.org/10.1016/j.apgeochem.2006.02.003} {\bibfield  {journal} {\bibinfo  {journal} {Applied Geochemistry}\ }\bibinfo {series} {Frontiers in {Analytical} {Geochemistry}–{An} {IGC} 2004 {Perspective}},\ \textbf {\bibinfo {volume} {21}},\ \bibinfo {pages} {730} (\bibinfo {year} {2006})}\BibitemShut
  {NoStop}%
\bibitem [{\citenamefont {DeLucia}\ \emph {et~al.}(2005)\citenamefont {DeLucia}, \citenamefont {Samuels}, \citenamefont {Harmon}, \citenamefont {Walters}, \citenamefont {McNesby}, \citenamefont {LaPointe}, \citenamefont {Winkel},\ and\ \citenamefont {Miziolek}}]{delucia_laser-induced_2005}%
  \BibitemOpen
  \bibfield  {author} {\bibinfo {author} {\bibfnamefont {F.}~\bibnamefont {DeLucia}}, \bibinfo {author} {\bibfnamefont {A.}~\bibnamefont {Samuels}}, \bibinfo {author} {\bibfnamefont {R.}~\bibnamefont {Harmon}}, \bibinfo {author} {\bibfnamefont {R.}~\bibnamefont {Walters}}, \bibinfo {author} {\bibfnamefont {K.}~\bibnamefont {McNesby}}, \bibinfo {author} {\bibfnamefont {A.}~\bibnamefont {LaPointe}}, \bibinfo {author} {\bibfnamefont {R.}~\bibnamefont {Winkel}},\ and\ \bibinfo {author} {\bibfnamefont {A.}~\bibnamefont {Miziolek}},\ }\bibfield  {title} {\bibinfo {title} {Laser-induced breakdown spectroscopy ({LIBS}): a promising versatile chemical sensor technology for hazardous material detection},\ }\href {https://doi.org/10.1109/JSEN.2005.848151} {\bibfield  {journal} {\bibinfo  {journal} {IEEE Sensors Journal}\ }\textbf {\bibinfo {volume} {5}},\ \bibinfo {pages} {681} (\bibinfo {year} {2005})}\BibitemShut {NoStop}%
\bibitem [{\citenamefont {Wysocki}\ and\ \citenamefont {Weidmann}(2010)}]{wysocki_molecular_2010}%
  \BibitemOpen
  \bibfield  {author} {\bibinfo {author} {\bibfnamefont {G.}~\bibnamefont {Wysocki}}\ and\ \bibinfo {author} {\bibfnamefont {D.}~\bibnamefont {Weidmann}},\ }\bibfield  {title} {\bibinfo {title} {Molecular dispersion spectroscopy for chemical sensing using chirped mid-infrared quantum cascade laser},\ }\href {https://doi.org/10.1364/OE.18.026123} {\bibfield  {journal} {\bibinfo  {journal} {Opt. Express, OE}\ }\textbf {\bibinfo {volume} {18}},\ \bibinfo {pages} {26123} (\bibinfo {year} {2010})}\BibitemShut {NoStop}%
\bibitem [{\citenamefont {Huet}\ \emph {et~al.}(2015)\citenamefont {Huet}, \citenamefont {Pettens},\ and\ \citenamefont {Bastin}}]{huet_isotope_2015}%
  \BibitemOpen
  \bibfield  {author} {\bibinfo {author} {\bibfnamefont {N.}~\bibnamefont {Huet}}, \bibinfo {author} {\bibfnamefont {M.}~\bibnamefont {Pettens}},\ and\ \bibinfo {author} {\bibfnamefont {T.}~\bibnamefont {Bastin}},\ }\bibfield  {title} {\bibinfo {title} {Isotope shifts and hyperfine structure of the laser-cooling {Fe} {I} 358-nm line},\ }\href {https://doi.org/10.1103/PhysRevA.92.052507} {\bibfield  {journal} {\bibinfo  {journal} {Phys. Rev. A}\ }\textbf {\bibinfo {volume} {92}},\ \bibinfo {pages} {052507} (\bibinfo {year} {2015})}\BibitemShut {NoStop}%
\bibitem [{\citenamefont {Phillips}(1998)}]{phillips_nobel_1998}%
  \BibitemOpen
  \bibfield  {author} {\bibinfo {author} {\bibfnamefont {W.~D.}\ \bibnamefont {Phillips}},\ }\bibfield  {title} {\bibinfo {title} {Nobel {Lecture}: {Laser} cooling and trapping of neutral atoms},\ }\href {https://doi.org/10.1103/RevModPhys.70.721} {\bibfield  {journal} {\bibinfo  {journal} {Rev. Mod. Phys.}\ }\textbf {\bibinfo {volume} {70}},\ \bibinfo {pages} {721} (\bibinfo {year} {1998})}\BibitemShut {NoStop}%
\bibitem [{\citenamefont {Degen}\ \emph {et~al.}(2017)\citenamefont {Degen}, \citenamefont {Reinhard},\ and\ \citenamefont {Cappellaro}}]{degen_quantum_2017}%
  \BibitemOpen
  \bibfield  {author} {\bibinfo {author} {\bibfnamefont {C.}~\bibnamefont {Degen}}, \bibinfo {author} {\bibfnamefont {F.}~\bibnamefont {Reinhard}},\ and\ \bibinfo {author} {\bibfnamefont {P.}~\bibnamefont {Cappellaro}},\ }\bibfield  {title} {\bibinfo {title} {Quantum sensing},\ }\href {https://doi.org/10.1103/RevModPhys.89.035002} {\bibfield  {journal} {\bibinfo  {journal} {Rev. Mod. Phys.}\ }\textbf {\bibinfo {volume} {89}},\ \bibinfo {pages} {035002} (\bibinfo {year} {2017})}\BibitemShut {NoStop}%
\bibitem [{\citenamefont {Anderson}\ \emph {et~al.}(1995)\citenamefont {Anderson}, \citenamefont {Ensher}, \citenamefont {Matthews}, \citenamefont {Wieman},\ and\ \citenamefont {Cornell}}]{anderson_observation_1995}%
  \BibitemOpen
  \bibfield  {author} {\bibinfo {author} {\bibfnamefont {M.~H.}\ \bibnamefont {Anderson}}, \bibinfo {author} {\bibfnamefont {J.~R.}\ \bibnamefont {Ensher}}, \bibinfo {author} {\bibfnamefont {M.~R.}\ \bibnamefont {Matthews}}, \bibinfo {author} {\bibfnamefont {C.~E.}\ \bibnamefont {Wieman}},\ and\ \bibinfo {author} {\bibfnamefont {E.~A.}\ \bibnamefont {Cornell}},\ }\bibfield  {title} {\bibinfo {title} {Observation of {Bose}-{Einstein} {Condensation} in a {Dilute} {Atomic} {Vapor}},\ }\href {https://doi.org/10.1126/science.269.5221.198} {\bibfield  {journal} {\bibinfo  {journal} {Science}\ }\textbf {\bibinfo {volume} {269}},\ \bibinfo {pages} {198} (\bibinfo {year} {1995})}\BibitemShut {NoStop}%
\bibitem [{\citenamefont {Prodan}\ and\ \citenamefont {Phillips}(1984)}]{prodan_chirping_1984}%
  \BibitemOpen
  \bibfield  {author} {\bibinfo {author} {\bibfnamefont {J.~V.}\ \bibnamefont {Prodan}}\ and\ \bibinfo {author} {\bibfnamefont {W.~D.}\ \bibnamefont {Phillips}},\ }\bibfield  {title} {\bibinfo {title} {Chirping the light—fantastic? {Recent} {NBS} atom cooling experiments},\ }\href {https://doi.org/10.1016/0079-6727(84)90019-3} {\bibfield  {journal} {\bibinfo  {journal} {Progress in Quantum Electronics}\ }\textbf {\bibinfo {volume} {8}},\ \bibinfo {pages} {231} (\bibinfo {year} {1984})}\BibitemShut {NoStop}%
\bibitem [{\citenamefont {Hutzler}\ \emph {et~al.}(2012)\citenamefont {Hutzler}, \citenamefont {Lu},\ and\ \citenamefont {Doyle}}]{hutzler_buffer_2012}%
  \BibitemOpen
  \bibfield  {author} {\bibinfo {author} {\bibfnamefont {N.~R.}\ \bibnamefont {Hutzler}}, \bibinfo {author} {\bibfnamefont {H.-I.}\ \bibnamefont {Lu}},\ and\ \bibinfo {author} {\bibfnamefont {J.~M.}\ \bibnamefont {Doyle}},\ }\bibfield  {title} {\bibinfo {title} {The {Buffer} {Gas} {Beam}: {An} {Intense}, {Cold}, and {Slow} {Source} for {Atoms} and {Molecules}},\ }\href {https://doi.org/10.1021/cr200362u} {\bibfield  {journal} {\bibinfo  {journal} {Chem. Rev.}\ }\textbf {\bibinfo {volume} {112}},\ \bibinfo {pages} {4803} (\bibinfo {year} {2012})}\BibitemShut {NoStop}%
\bibitem [{\citenamefont {Kock}\ \emph {et~al.}(2016)\citenamefont {Kock}, \citenamefont {He}, \citenamefont {{\'S}wierad}, \citenamefont {Smith}, \citenamefont {Hughes}, \citenamefont {Bongs},\ and\ \citenamefont {Singh}}]{kock2016laser}%
  \BibitemOpen
  \bibfield  {author} {\bibinfo {author} {\bibfnamefont {O.}~\bibnamefont {Kock}}, \bibinfo {author} {\bibfnamefont {W.}~\bibnamefont {He}}, \bibinfo {author} {\bibfnamefont {D.}~\bibnamefont {{\'S}wierad}}, \bibinfo {author} {\bibfnamefont {L.}~\bibnamefont {Smith}}, \bibinfo {author} {\bibfnamefont {J.}~\bibnamefont {Hughes}}, \bibinfo {author} {\bibfnamefont {K.}~\bibnamefont {Bongs}},\ and\ \bibinfo {author} {\bibfnamefont {Y.}~\bibnamefont {Singh}},\ }\bibfield  {title} {\bibinfo {title} {Laser controlled atom source for optical clocks},\ }\href@noop {} {\bibfield  {journal} {\bibinfo  {journal} {Scientific reports}\ }\textbf {\bibinfo {volume} {6}},\ \bibinfo {pages} {37321} (\bibinfo {year} {2016})}\BibitemShut {NoStop}%
\bibitem [{\citenamefont {Yasuda}\ \emph {et~al.}(2017)\citenamefont {Yasuda}, \citenamefont {Tanabe}, \citenamefont {Kobayashi}, \citenamefont {Akamatsu}, \citenamefont {Sato},\ and\ \citenamefont {Hatakeyama}}]{yasuda2017laser}%
  \BibitemOpen
  \bibfield  {author} {\bibinfo {author} {\bibfnamefont {M.}~\bibnamefont {Yasuda}}, \bibinfo {author} {\bibfnamefont {T.}~\bibnamefont {Tanabe}}, \bibinfo {author} {\bibfnamefont {T.}~\bibnamefont {Kobayashi}}, \bibinfo {author} {\bibfnamefont {D.}~\bibnamefont {Akamatsu}}, \bibinfo {author} {\bibfnamefont {T.}~\bibnamefont {Sato}},\ and\ \bibinfo {author} {\bibfnamefont {A.}~\bibnamefont {Hatakeyama}},\ }\bibfield  {title} {\bibinfo {title} {Laser-controlled cold ytterbium atom source for transportable optical clocks},\ }\href@noop {} {\bibfield  {journal} {\bibinfo  {journal} {Journal of the Physical Society of Japan}\ }\textbf {\bibinfo {volume} {86}},\ \bibinfo {pages} {125001} (\bibinfo {year} {2017})}\BibitemShut {NoStop}%
\bibitem [{\citenamefont {Stephens}\ \emph {et~al.}(1994)\citenamefont {Stephens}, \citenamefont {Rhodes},\ and\ \citenamefont {Wieman}}]{stephens_study_1994}%
  \BibitemOpen
  \bibfield  {author} {\bibinfo {author} {\bibfnamefont {M.}~\bibnamefont {Stephens}}, \bibinfo {author} {\bibfnamefont {R.}~\bibnamefont {Rhodes}},\ and\ \bibinfo {author} {\bibfnamefont {C.}~\bibnamefont {Wieman}},\ }\bibfield  {title} {\bibinfo {title} {Study of wall coatings for vapor-cell laser traps},\ }\href {https://doi.org/10.1063/1.358502} {\bibfield  {journal} {\bibinfo  {journal} {Journal of Applied Physics}\ }\textbf {\bibinfo {volume} {76}},\ \bibinfo {pages} {3479} (\bibinfo {year} {1994})}\BibitemShut {NoStop}%
\bibitem [{\citenamefont {Goldenberg}\ \emph {et~al.}(1961)\citenamefont {Goldenberg}, \citenamefont {Kleppner},\ and\ \citenamefont {Ramsey}}]{goldenberg_atomic_1961}%
  \BibitemOpen
  \bibfield  {author} {\bibinfo {author} {\bibfnamefont {H.~M.}\ \bibnamefont {Goldenberg}}, \bibinfo {author} {\bibfnamefont {D.}~\bibnamefont {Kleppner}},\ and\ \bibinfo {author} {\bibfnamefont {N.~F.}\ \bibnamefont {Ramsey}},\ }\bibfield  {title} {\bibinfo {title} {Atomic {Beam} {Resonance} {Experiments} with {Stored} {Beams}},\ }\href {https://doi.org/10.1103/PhysRev.123.530} {\bibfield  {journal} {\bibinfo  {journal} {Phys. Rev.}\ }\textbf {\bibinfo {volume} {123}},\ \bibinfo {pages} {530} (\bibinfo {year} {1961})}\BibitemShut {NoStop}%
\bibitem [{\citenamefont {Rahman}\ and\ \citenamefont {Robinson}(1987)}]{rahman1987rb}%
  \BibitemOpen
  \bibfield  {author} {\bibinfo {author} {\bibfnamefont {C.}~\bibnamefont {Rahman}}\ and\ \bibinfo {author} {\bibfnamefont {H.}~\bibnamefont {Robinson}},\ }\bibfield  {title} {\bibinfo {title} {Rb oo hyperfine transition in evacuated wall-coated cell at melting temperature},\ }\href@noop {} {\bibfield  {journal} {\bibinfo  {journal} {IEEE journal of quantum electronics}\ }\textbf {\bibinfo {volume} {23}},\ \bibinfo {pages} {452} (\bibinfo {year} {1987})}\BibitemShut {NoStop}%
\bibitem [{\citenamefont {Yi}\ \emph {et~al.}(2008)\citenamefont {Yi}, \citenamefont {Robinson}, \citenamefont {Knappe}, \citenamefont {Maclennan}, \citenamefont {Jones}, \citenamefont {Zhu}, \citenamefont {Clark},\ and\ \citenamefont {Kitching}}]{yi2008method}%
  \BibitemOpen
  \bibfield  {author} {\bibinfo {author} {\bibfnamefont {Y.}~\bibnamefont {Yi}}, \bibinfo {author} {\bibfnamefont {H.}~\bibnamefont {Robinson}}, \bibinfo {author} {\bibfnamefont {S.}~\bibnamefont {Knappe}}, \bibinfo {author} {\bibfnamefont {J.}~\bibnamefont {Maclennan}}, \bibinfo {author} {\bibfnamefont {C.}~\bibnamefont {Jones}}, \bibinfo {author} {\bibfnamefont {C.}~\bibnamefont {Zhu}}, \bibinfo {author} {\bibfnamefont {N.}~\bibnamefont {Clark}},\ and\ \bibinfo {author} {\bibfnamefont {J.}~\bibnamefont {Kitching}},\ }\bibfield  {title} {\bibinfo {title} {Method for characterizing self-assembled monolayers as antirelaxation wall coatings for alkali vapor cells},\ }\href@noop {} {\bibfield  {journal} {\bibinfo  {journal} {Journal of Applied Physics}\ }\textbf {\bibinfo {volume} {104}} (\bibinfo {year} {2008})}\BibitemShut {NoStop}%
\bibitem [{\citenamefont {Chi}\ \emph {et~al.}(2020)\citenamefont {Chi}, \citenamefont {Quan}, \citenamefont {Zhang}, \citenamefont {Zhao},\ and\ \citenamefont {Fang}}]{chi_advances_2020}%
  \BibitemOpen
  \bibfield  {author} {\bibinfo {author} {\bibfnamefont {H.}~\bibnamefont {Chi}}, \bibinfo {author} {\bibfnamefont {W.}~\bibnamefont {Quan}}, \bibinfo {author} {\bibfnamefont {J.}~\bibnamefont {Zhang}}, \bibinfo {author} {\bibfnamefont {L.}~\bibnamefont {Zhao}},\ and\ \bibinfo {author} {\bibfnamefont {J.}~\bibnamefont {Fang}},\ }\bibfield  {title} {\bibinfo {title} {Advances in anti-relaxation coatings of alkali-metal vapor cells},\ }\href {https://doi.org/10.1016/j.apsusc.2019.143897} {\bibfield  {journal} {\bibinfo  {journal} {Applied Surface Science}\ }\textbf {\bibinfo {volume} {501}},\ \bibinfo {pages} {143897} (\bibinfo {year} {2020})}\BibitemShut {NoStop}%
\bibitem [{\citenamefont {Sekiguchi}\ \emph {et~al.}(2018)\citenamefont {Sekiguchi}, \citenamefont {Hatakeyama}, \citenamefont {Okuma},\ and\ \citenamefont {Usui}}]{sekiguchi_scattering_2018}%
  \BibitemOpen
  \bibfield  {author} {\bibinfo {author} {\bibfnamefont {N.}~\bibnamefont {Sekiguchi}}, \bibinfo {author} {\bibfnamefont {A.}~\bibnamefont {Hatakeyama}}, \bibinfo {author} {\bibfnamefont {K.}~\bibnamefont {Okuma}},\ and\ \bibinfo {author} {\bibfnamefont {H.}~\bibnamefont {Usui}},\ }\bibfield  {title} {\bibinfo {title} {Scattering of an alkali-metal atomic beam on anti-spin-relaxation coatings},\ }\href {https://doi.org/10.1103/PhysRevA.98.042709} {\bibfield  {journal} {\bibinfo  {journal} {Phys. Rev. A}\ }\textbf {\bibinfo {volume} {98}},\ \bibinfo {pages} {042709} (\bibinfo {year} {2018})}\BibitemShut {NoStop}%
\bibitem [{\citenamefont {Wu}(2021)}]{wu_wall_2021}%
  \BibitemOpen
  \bibfield  {author} {\bibinfo {author} {\bibfnamefont {Z.}~\bibnamefont {Wu}},\ }\bibfield  {title} {\bibinfo {title} {Wall interactions of spin-polarized atoms},\ }\href {https://doi.org/10.1103/RevModPhys.93.035006} {\bibfield  {journal} {\bibinfo  {journal} {Rev. Mod. Phys.}\ }\textbf {\bibinfo {volume} {93}},\ \bibinfo {pages} {035006} (\bibinfo {year} {2021})}\BibitemShut {NoStop}%
\bibitem [{\citenamefont {Liberman}\ and\ \citenamefont {Knize}(1986)}]{liberman1986relaxation}%
  \BibitemOpen
  \bibfield  {author} {\bibinfo {author} {\bibfnamefont {V.}~\bibnamefont {Liberman}}\ and\ \bibinfo {author} {\bibfnamefont {R.}~\bibnamefont {Knize}},\ }\bibfield  {title} {\bibinfo {title} {Relaxation of optically pumped cs in wall-coated cells},\ }\href@noop {} {\bibfield  {journal} {\bibinfo  {journal} {Physical Review A}\ }\textbf {\bibinfo {volume} {34}},\ \bibinfo {pages} {5115} (\bibinfo {year} {1986})}\BibitemShut {NoStop}%
\bibitem [{\citenamefont {Budker}\ \emph {et~al.}(2005)\citenamefont {Budker}, \citenamefont {Hollberg}, \citenamefont {Kimball}, \citenamefont {Kitching}, \citenamefont {Pustelny},\ and\ \citenamefont {Yashchuk}}]{budker2005microwave}%
  \BibitemOpen
  \bibfield  {author} {\bibinfo {author} {\bibfnamefont {D.}~\bibnamefont {Budker}}, \bibinfo {author} {\bibfnamefont {L.}~\bibnamefont {Hollberg}}, \bibinfo {author} {\bibfnamefont {D.~F.}\ \bibnamefont {Kimball}}, \bibinfo {author} {\bibfnamefont {J.}~\bibnamefont {Kitching}}, \bibinfo {author} {\bibfnamefont {S.}~\bibnamefont {Pustelny}},\ and\ \bibinfo {author} {\bibfnamefont {V.~V.}\ \bibnamefont {Yashchuk}},\ }\bibfield  {title} {\bibinfo {title} {Microwave transitions and nonlinear magneto-optical rotation in anti-relaxation-coated cells},\ }\href@noop {} {\bibfield  {journal} {\bibinfo  {journal} {Physical Review A—Atomic, Molecular, and Optical Physics}\ }\textbf {\bibinfo {volume} {71}},\ \bibinfo {pages} {012903} (\bibinfo {year} {2005})}\BibitemShut {NoStop}%
\bibitem [{\citenamefont {Atutov}\ and\ \citenamefont {Plekhanov}(2015)}]{atutov_accurate_2015}%
  \BibitemOpen
  \bibfield  {author} {\bibinfo {author} {\bibfnamefont {S.~N.}\ \bibnamefont {Atutov}}\ and\ \bibinfo {author} {\bibfnamefont {A.~I.}\ \bibnamefont {Plekhanov}},\ }\bibfield  {title} {\bibinfo {title} {Accurate measurement of the sticking time and sticking probability of {Rb} atoms on a polydimethylsiloxane coating},\ }\href {https://doi.org/10.1134/S1063776115010094} {\bibfield  {journal} {\bibinfo  {journal} {J. Exp. Theor. Phys.}\ }\textbf {\bibinfo {volume} {120}},\ \bibinfo {pages} {1} (\bibinfo {year} {2015})}\BibitemShut {NoStop}%
\bibitem [{\citenamefont {Bouchiat}\ and\ \citenamefont {Brossel}(1966)}]{bouchiat_relaxation_1966}%
  \BibitemOpen
  \bibfield  {author} {\bibinfo {author} {\bibfnamefont {M.~A.}\ \bibnamefont {Bouchiat}}\ and\ \bibinfo {author} {\bibfnamefont {J.}~\bibnamefont {Brossel}},\ }\bibfield  {title} {\bibinfo {title} {Relaxation of {Optically} {Pumped} {Rb} {Atoms} on {Paraffin}-{Coated} {Walls}},\ }\href {https://doi.org/10.1103/PhysRev.147.41} {\bibfield  {journal} {\bibinfo  {journal} {Phys. Rev.}\ }\textbf {\bibinfo {volume} {147}},\ \bibinfo {pages} {41} (\bibinfo {year} {1966})}\BibitemShut {NoStop}%
\bibitem [{\citenamefont {Zhao}\ \emph {et~al.}(2009)\citenamefont {Zhao}, \citenamefont {Schaden},\ and\ \citenamefont {Wu}}]{zhao_method_2009}%
  \BibitemOpen
  \bibfield  {author} {\bibinfo {author} {\bibfnamefont {K.~F.}\ \bibnamefont {Zhao}}, \bibinfo {author} {\bibfnamefont {M.}~\bibnamefont {Schaden}},\ and\ \bibinfo {author} {\bibfnamefont {Z.}~\bibnamefont {Wu}},\ }\bibfield  {title} {\bibinfo {title} {Method for {Measuring} the {Dwell} {Time} of {Spin}-{Polarized} {Rb} {Atoms} on {Coated} {Pyrex} {Glass} {Surfaces} {Using} {Light} {Shift}},\ }\href {https://doi.org/10.1103/PhysRevLett.103.073201} {\bibfield  {journal} {\bibinfo  {journal} {Phys. Rev. Lett.}\ }\textbf {\bibinfo {volume} {103}},\ \bibinfo {pages} {073201} (\bibinfo {year} {2009})}\BibitemShut {NoStop}%
\bibitem [{\citenamefont {Ulanski}\ and\ \citenamefont {Wu}(2011)}]{ulanski_measurement_2011}%
  \BibitemOpen
  \bibfield  {author} {\bibinfo {author} {\bibfnamefont {E.}~\bibnamefont {Ulanski}}\ and\ \bibinfo {author} {\bibfnamefont {Z.}~\bibnamefont {Wu}},\ }\bibfield  {title} {\bibinfo {title} {Measurement of dwell times of spin polarized rubidium atoms on octadecyltrichlorosilane- and paraffin-coated surfaces},\ }\href {https://doi.org/10.1063/1.3591172} {\bibfield  {journal} {\bibinfo  {journal} {Applied Physics Letters}\ }\textbf {\bibinfo {volume} {98}},\ \bibinfo {pages} {201115} (\bibinfo {year} {2011})}\BibitemShut {NoStop}%
\bibitem [{\citenamefont {Gozzini}\ \emph {et~al.}(1993)\citenamefont {Gozzini}, \citenamefont {Mango}, \citenamefont {Xu}, \citenamefont {Alzetta}, \citenamefont {Maccarrone},\ and\ \citenamefont {Bernheim}}]{gozzini_light-induced_1993}%
  \BibitemOpen
  \bibfield  {author} {\bibinfo {author} {\bibfnamefont {A.}~\bibnamefont {Gozzini}}, \bibinfo {author} {\bibfnamefont {F.}~\bibnamefont {Mango}}, \bibinfo {author} {\bibfnamefont {J.~H.}\ \bibnamefont {Xu}}, \bibinfo {author} {\bibfnamefont {G.}~\bibnamefont {Alzetta}}, \bibinfo {author} {\bibfnamefont {F.}~\bibnamefont {Maccarrone}},\ and\ \bibinfo {author} {\bibfnamefont {R.~A.}\ \bibnamefont {Bernheim}},\ }\bibfield  {title} {\bibinfo {title} {Light-induced ejection of alkali atoms in polysiloxane coated cells},\ }\href {https://doi.org/10.1007/BF02482437} {\bibfield  {journal} {\bibinfo  {journal} {Il Nuovo Cimento D}\ }\textbf {\bibinfo {volume} {15}},\ \bibinfo {pages} {709} (\bibinfo {year} {1993})}\BibitemShut {NoStop}%
\bibitem [{\citenamefont {Atutov}\ \emph {et~al.}(1999)\citenamefont {Atutov}, \citenamefont {Biancalana}, \citenamefont {Bicchi}, \citenamefont {Marinelli}, \citenamefont {Mariotti}, \citenamefont {Meucci}, \citenamefont {Nagel}, \citenamefont {Nasyrov}, \citenamefont {Rachini},\ and\ \citenamefont {Moi}}]{atutov_light-induced_1999}%
  \BibitemOpen
  \bibfield  {author} {\bibinfo {author} {\bibfnamefont {S.~N.}\ \bibnamefont {Atutov}}, \bibinfo {author} {\bibfnamefont {V.}~\bibnamefont {Biancalana}}, \bibinfo {author} {\bibfnamefont {P.}~\bibnamefont {Bicchi}}, \bibinfo {author} {\bibfnamefont {C.}~\bibnamefont {Marinelli}}, \bibinfo {author} {\bibfnamefont {E.}~\bibnamefont {Mariotti}}, \bibinfo {author} {\bibfnamefont {M.}~\bibnamefont {Meucci}}, \bibinfo {author} {\bibfnamefont {A.}~\bibnamefont {Nagel}}, \bibinfo {author} {\bibfnamefont {K.~A.}\ \bibnamefont {Nasyrov}}, \bibinfo {author} {\bibfnamefont {S.}~\bibnamefont {Rachini}},\ and\ \bibinfo {author} {\bibfnamefont {L.}~\bibnamefont {Moi}},\ }\bibfield  {title} {\bibinfo {title} {Light-induced diffusion and desorption of alkali metals in a siloxane film: {Theory} and experiment},\ }\href {https://doi.org/10.1103/PhysRevA.60.4693} {\bibfield  {journal} {\bibinfo  {journal} {Phys. Rev. A}\ }\textbf {\bibinfo {volume} {60}},\ \bibinfo {pages} {4693} (\bibinfo {year} {1999})}\BibitemShut {NoStop}%
\bibitem [{\citenamefont {Asakawa}\ \emph {et~al.}(2021)\citenamefont {Asakawa}, \citenamefont {Tanaka}, \citenamefont {Uemura}, \citenamefont {Matsuzaka}, \citenamefont {Nishikawa}, \citenamefont {Oguma}, \citenamefont {Usui},\ and\ \citenamefont {Hatakeyama}}]{asakawa_measurement_2021}%
  \BibitemOpen
  \bibfield  {author} {\bibinfo {author} {\bibfnamefont {K.}~\bibnamefont {Asakawa}}, \bibinfo {author} {\bibfnamefont {Y.}~\bibnamefont {Tanaka}}, \bibinfo {author} {\bibfnamefont {K.}~\bibnamefont {Uemura}}, \bibinfo {author} {\bibfnamefont {N.}~\bibnamefont {Matsuzaka}}, \bibinfo {author} {\bibfnamefont {K.}~\bibnamefont {Nishikawa}}, \bibinfo {author} {\bibfnamefont {Y.}~\bibnamefont {Oguma}}, \bibinfo {author} {\bibfnamefont {H.}~\bibnamefont {Usui}},\ and\ \bibinfo {author} {\bibfnamefont {A.}~\bibnamefont {Hatakeyama}},\ }\bibfield  {title} {\bibinfo {title} {Measurement of the temperature dependence of dwell time and spin relaxation probability of {Rb} atoms on paraffin surfaces using a beam-scattering method},\ }\href {https://doi.org/10.1103/PhysRevA.104.063106} {\bibfield  {journal} {\bibinfo  {journal} {Phys. Rev. A}\ }\textbf {\bibinfo {volume} {104}},\ \bibinfo {pages} {063106} (\bibinfo {year} {2021})}\BibitemShut {NoStop}%
\bibitem [{sup()}]{supplemental}%
  \BibitemOpen
  \href@noop {} {\bibinfo {title} {See supplemental material at \text{[URL will be inserted by publisher]} for an optical image of the spin-coated \text{PDMS} surface(\text{Fig. S 1}) and more details of the numerical simulation showing the flux enchancement}}\BibitemShut {NoStop}%
\bibitem [{\citenamefont {Hall}\ \emph {et~al.}(1998)\citenamefont {Hall}, \citenamefont {Underhill},\ and\ \citenamefont {Torkelson}}]{hall_spin_1998}%
  \BibitemOpen
  \bibfield  {author} {\bibinfo {author} {\bibfnamefont {D.~B.}\ \bibnamefont {Hall}}, \bibinfo {author} {\bibfnamefont {P.}~\bibnamefont {Underhill}},\ and\ \bibinfo {author} {\bibfnamefont {J.~M.}\ \bibnamefont {Torkelson}},\ }\bibfield  {title} {\bibinfo {title} {Spin coating of thin and ultrathin polymer films},\ }\href {https://doi.org/10.1002/pen.10373} {\bibfield  {journal} {\bibinfo  {journal} {Polymer Engineering \& Science}\ }\textbf {\bibinfo {volume} {38}},\ \bibinfo {pages} {2039} (\bibinfo {year} {1998})}\BibitemShut {NoStop}%
\bibitem [{\citenamefont {Bračič}\ \emph {et~al.}(2014)\citenamefont {Bračič}, \citenamefont {Mohan}, \citenamefont {Kargl}, \citenamefont {Griesser}, \citenamefont {Hribernik}, \citenamefont {Köstler}, \citenamefont {Stana-Kleinschek},\ and\ \citenamefont {Fras-Zemljič}}]{bracic_preparation_2014}%
  \BibitemOpen
  \bibfield  {author} {\bibinfo {author} {\bibfnamefont {M.}~\bibnamefont {Bračič}}, \bibinfo {author} {\bibfnamefont {T.}~\bibnamefont {Mohan}}, \bibinfo {author} {\bibfnamefont {R.}~\bibnamefont {Kargl}}, \bibinfo {author} {\bibfnamefont {T.}~\bibnamefont {Griesser}}, \bibinfo {author} {\bibfnamefont {S.}~\bibnamefont {Hribernik}}, \bibinfo {author} {\bibfnamefont {S.}~\bibnamefont {Köstler}}, \bibinfo {author} {\bibfnamefont {K.}~\bibnamefont {Stana-Kleinschek}},\ and\ \bibinfo {author} {\bibfnamefont {L.}~\bibnamefont {Fras-Zemljič}},\ }\bibfield  {title} {\bibinfo {title} {Preparation of {PDMS} ultrathin films and patterned surface modification with cellulose},\ }\href {https://doi.org/10.1039/C3RA47380E} {\bibfield  {journal} {\bibinfo  {journal} {RSC Advances}\ }\textbf {\bibinfo {volume} {4}},\ \bibinfo {pages} {11955} (\bibinfo {year} {2014})}\BibitemShut {NoStop}%
\bibitem [{\citenamefont {Bucay}(2019)}]{bucay_surface_2019}%
  \BibitemOpen
  \bibfield  {author} {\bibinfo {author} {\bibfnamefont {I.}~\bibnamefont {Bucay}},\ }\emph {\bibinfo {title} {Surface {Ionization} of {Metastable} {Calcium} and {Ytterbium} {Atoms}}},\ \href {https://www.proquest.com/docview/2479389916/abstract/7C2F16124BA342D8PQ/1} {\bibinfo {type} {Ph.{D}.}},\ \bibinfo  {school} {The University of Texas at Austin}, \bibinfo {address} {United States -- Texas} (\bibinfo {year} {2019}),\ \bibinfo {note} {iSBN: 9798684608384}\BibitemShut {NoStop}%
\bibitem [{\citenamefont {Zhang}\ \emph {et~al.}(2020)\citenamefont {Zhang}, \citenamefont {Sun}, \citenamefont {Qian}, \citenamefont {Gao},\ and\ \citenamefont {Zuo}}]{zhang_experimental_2020}%
  \BibitemOpen
  \bibfield  {author} {\bibinfo {author} {\bibfnamefont {G.}~\bibnamefont {Zhang}}, \bibinfo {author} {\bibfnamefont {Y.}~\bibnamefont {Sun}}, \bibinfo {author} {\bibfnamefont {B.}~\bibnamefont {Qian}}, \bibinfo {author} {\bibfnamefont {H.}~\bibnamefont {Gao}},\ and\ \bibinfo {author} {\bibfnamefont {D.}~\bibnamefont {Zuo}},\ }\bibfield  {title} {\bibinfo {title} {Experimental study on mechanical performance of polydimethylsiloxane ({PDMS}) at various temperatures},\ }\href {https://doi.org/10.1016/j.polymertesting.2020.106670} {\bibfield  {journal} {\bibinfo  {journal} {Polymer Testing}\ }\textbf {\bibinfo {volume} {90}},\ \bibinfo {pages} {106670} (\bibinfo {year} {2020})}\BibitemShut {NoStop}%
\end{thebibliography}%

\end{document}